
This paper refers to seven figures (not included). Hard copies of
the figures will be mailed upon request.
\input phyzzx
%
%
\catcode`\@=11 
\def\papers{\papersize\headline=\paperheadline\footline=\paperfootline}
\def\papersize{\hsize=40pc \vsize=53pc \hoffset=0pc \voffset=1pc
   \advance\hoffset by\HOFFSET \advance\voffset by\VOFFSET
   \pagebottomfiller=0pc
   \skip\footins=\bigskipamount \normalspace }
\catcode`\@=12 
\papers

\def\square{\kern1pt\vbox{\hrule height 1.2pt\hbox{\vrule width 1.2pt\hskip 3pt
   \vbox{\vskip 6pt}\hskip 3pt\vrule width 0.6pt}\hrule height 0.6pt}\kern1pt}

\def\L{{\cal L}}

\def\V{{\cal V}}

\baselineskip 13pt plus 1pt minus 1pt
\overfullrule=0pt
\pubnum{IASSNS-HEP-92/8 \cr
MIT-CTP-2067}
\date{February 1992}
\titlepage
\title{INTERPOLATING STRING FIELD THEORIES}
\author{Barton Zwiebach
\foot{Permanent address: Center for Theoretical Physics, MIT,
Cambridge, Mass. 02139. Supported in part by D.O.E. contract
DE-AC02-76ER03069 and NSF grant PHY91-06210.}}
\address{School of Natural Sciences \break
Institute for Advanced Study \break
Olden Lane \break
Princeton, NJ 08540}
\abstract{ A minimal area problem imposing different length conditions
on open and closed curves is shown to define a one parameter family of
covariant open-closed quantum string field theories. These
interpolate from a recently proposed factorizable open-closed
theory up to an extended version of Witten's open string field
theory capable of incorporating on shell closed strings.
The string diagrams of the latter define a new decomposition
of the moduli spaces of Riemann surfaces with punctures and
boundaries based on quadratic differentials
with both first order and second order poles.}
\endpage

\REF\zwiebach{B. Zwiebach, ``Quantum Open String Theory with Manifest
Closed String Factorization'', Phys. Lett. {\bf B256} (1991) 22.}
\REF\witten{E. Witten, ``Noncommutative Geometry and String Field
Theory'', Nucl. Phys. {\bf B268} (1986) 253.}
\REF\zwiebachcmp{B. Zwiebach, `` A proof that Witten's open string
theory gives a single cover of moduli space'',
Comm. Math. Phys.}
\REF\shapirothorn{J. A. Shapiro and C. B. Thorn, Phys. Lett. {\bf B194}
(1987) 43.}
\REF\strominger{A. Strominger, Phys. Rev. Lett. {\bf 58} (1987) 58.}
\REF\batalinvilkovisky{I. A. Batalin and G. A. Vilkovisky, Phys.
Rev. {\bf D28} (1983) 2567.}
\REF\zwiebachqcs{B. Zwiebach, Mod. Phys. Lett. A5 (1990) 2753.}
\REF\zwiebachmam{B. Zwiebach, Comm. Math. Phys. {\bf 136} (1991) 83.}
\REF\strebel{K. Strebel, {\it Quadratic Differentials},
Springer-Verlag, Berlin, 1984.}
\REF\saadizwiebach{M. Saadi and B. Zwiebach, Ann. Phys. 192 (1989) 213.}
\REF\sen{A. Sen, ``On the Background Independence of String Field
Theory. III. Field Redefinitions'', Tata Institute Preprint, TIFR-TH-92/03,
January 1992.}
\REF\kontsevich{M. Kontsevich, ``Intersection Theory in the Moduli
Space of Curves and the Matrix Airy Function'', Max-Planck-Institut
preprint, MPI/91-77.}
\REF\thorn{C. B. Thorn, ``String field theory'', Phys.
Rep. {\bf 174} (1989) 1.}
\REF\sonodazwiebach{H. Sonoda and B. Zwiebach, Nucl. Phys.
{\bf B331} (1990) 592.}
\REF\zwiebachtalk{B. Zwiebach, `` Recursion Relations in Closed String
Field Theory'', Proceedings of the ``Strings 90'' Superstring
Workshop at Texas A\& M. Eds. R. Arnowitt, et. al.
(World Scientific, 1991) pp.266-275.}
\REF\zwiebachcmpnext{B. Zwiebach,`` Minimal area problems
and quantum open strings'', Comm. Math. Phys. {\bf 141} (1991) 577.}

A covariant quantum theory of open and closed strings was
derived recently from string diagrams defined by
a minimal area problem [\zwiebach ]. This problem required that the
length of any nontrivial open curve be larger or equal to $l_o =\pi$,
and that the length of any nontrivial closed curve be larger or equal
to  $l_c=2\pi$.
In the resulting theory one can compute directly all scattering
amplitudes involving any possible numbers of open and closed strings on
any surface. The Feynman rules of covariant open string theory [\witten]
are simpler, but amplitudes
involving external closed strings must be obtained indirectly
by factorization. Finding a way of extending the simple string diagrams
of open string theory to include external closed strings was the
motivation for the present work.

The solution is simple. Open string diagrams are now known
to arise from minimal area metrics
under the condition that all nontrivial open curves
be longer or equal to $\pi$ [\zwiebachcmp ]. We can incorporate external
closed strings without changing the minimal area problem at all!$\,$
Closed string external states correspond to punctures inside the
surface, and curves cannot be moved across them. Their
presence creates new homotopy types of nontrivial open curves.
The minimal area metric will have to adjust itself in order to satisfy
the new length conditions arising due to the closed string punctures.

The above minimal area problem is recognized to correspond to
$l_o =\pi$, and $l_c=0$. It is therefore natural to consider the
generalized problem where $l_o=\pi$ and $l_c=a$, where $a$ is
a constant. We shall see that this defines a one parameter
family of string diagrams. Even more interesting is that actually
they correspond to a one parameter family of string field theories.
This happens simply because the string diagrams can be built by
sewing. When $a=2\pi$ we recover the open-closed
factorizable theory [\zwiebach ], and when $a=0$ we get the extended open
string theory.

A large part of our effort in this paper will be devoted to
elucidate the extended open string theory. Since the minimal area
problem is showing us a natural way to incorporate external closed
strings into the framework of the open string theory we only
have to figure out how the string diagrams look, and if they can
be built using Feynman rules.
The minimal area metrics we shall find always arise from quadratic
differentials, and closed strings appear as first order poles.
A first order pole corresponds to a finite area conical
singularity of the metric with an angle of $\pi$ at the singularity.
We will find the amazingly simple result that a single open-closed
interaction is all one needs to add to the Witten open string
vertex so that the Feynman rules generate the above minimal area
string diagrams and thus generate all moduli spaces relevant
for open and open-closed string scattering amplitudes. There is no
need to add even an open-open-closed vertex. The open-closed vertex
appeared first in the factorization studies of Shapiro and Thorn
[\shapirothorn ] and in the work of Strominger [\strominger ].
Its geometry corresponds to that of the identity field in open
string field theory [\witten].

Our work shows that this
vertex, together with the open string vertex, generate
a previously unsuspected decomposition of the moduli spaces of surfaces
with boundaries and any number of punctures on the boundary and
on the interior. The resulting Feynman rules
have limitations: the closed strings must be on-shell, and, one
cannot compute directly pure closed string processes.
This is the price we pay for the extraordinarily simple
decomposition of moduli space we achieve with the open-closed vertex.
(Such difficulties might be generic in polynomial formulations).
For a fully field-theoretical way of incorporating closed
strings in open string theory we must use Ref. [\zwiebach ].

For closed strings we can generalize the usual string diagrams
by imposing different length conditions on closed curves
that correspond to external or internal strings
(mathematically, on closed curves homotopic or nonhomotopic to
punctures respectively). The resulting string diagrams interpolate
between the standard covariant closed string diagrams [\zwiebachqcs,
\zwiebachmam ], the Strebel type string diagrams [\strebel,
\saadizwiebach] and a set of diagrams where the external closed strings
appear as conical singularities of finite area. Here it is not clear that
the family of string diagrams yield a family of string field theories.

The present results could give us some flexibility in dealing with
the issue of background independence of string field theory. The
reason is the following. After shifting by a classical solution, the
resulting string field theory around the new background is defined
in terms of a new set of effective vertices.
It is possible, but difficult, to show that
there is a field redefinition bringing the action back to the original
form [\sen ].  Perhaps the new effective vertices could be related
to alternative vertices giving simply another decomposition
of moduli space. At a more technical level, the reason why it is
not manifest that a shift of the string field, via the three string
vertex, yields a change in the BRST operator is that the latter is an
insertion of a local operator, while the (covariant) vertex does not
give a local insertion, it gives a smeared insertion, because
all three strings in the vertex have finite length. The possibility
of changing the length of a string consistently could help relate a
local insertion to a smeared insertion. The open-closed theory
for $l_c$ very small but finite may allow to treat the effects of
changes in closed string backgrounds as local insertions in open
string worldsheets. An alternative application might be in matrix models.
Since the cubic action of open string field theory and its associated
decomposition of moduli space are closely related
to a matrix model with a background matrix [\kontsevich ], the action
that will be discussed below, giving a new decomposition of moduli
space could lead to a related matrix model integral.
Finally, some physicists believe that the final form of string field
theory should incorporate all possible ways of decomposing moduli space.
Understanding how a family of decompositions (such as that presented
here) could be put together into a single theory would be a first step
in such program.

Before we begin, we should comment on the rigor of the presentation.
We will prove that the string diagrams
including closed strings are solutions
of the minimal area problem, thus showing there cannot be overcounting.
We will not, however, give an existence theorem for these quadratic
differentials, so in principle, the Feynman rules could leave holes
in moduli space. This seems very unlikely to be the case, because
we will show, by analyzing recursion relations, that except for
singular surfaces, the boundaries of the Feynman graphs contibuting
to any amplitude always match.
\medskip
\noindent
$\underline{\hbox{The Open-Closed Vertex}}.\,$ Here we are looking
for a string diagram for a disk containing one open and one closed
string puncture. On the string diagram any nontrivial
open curve should be longer than $\pi$, and the area ought to
be minimal. If we had an infinite strip it would correspond to a
disk with two open string punctures, thus the strip must be
semiinfinite, and of width $\pi$, as shown in Fig.~1(a). Let us
find out how to place the closed string puncture on this strip.
Suppose it is located at $A$, namely at a point off the line
cutting the strip in two equal pieces (shown with dots in the figure).
This leads to a contradiction since there
would be nontrivial open curves shorter than $\pi$. The puncture
must therefore be on the middle of the strip.
Suppose the puncture is at $B$ (on the middle line but away from
the end of the semiinfinite strip). This is still
problematic. It contradicts the saturation condition [\zwiebachqcs ]
which requires that through {\it every} point $Q$ on the surface there
should be a nontrivial open curve of length $\pi$.\footnote{*}{If this
is not the case, the value of the metric on a small neighborhood of
$Q$ can be reduced
without violating any length condition, and thus the area can be reduced.
This contradicts the fact that the metric is supposed to be of minimal area.}
There is, however, no such curve going through point $Q$ (Fig.~1(a)),
the obvious candidate, a vertical segment
through $Q$ is homotopically trivial.
This shows the puncture should be
pushed all the way to $C$ (the end of the strip). There, however,
it would lie on the boundary of the strip, which is not possible.
We must therefore identify the $CD$ and $CD'$ segments.
The end result is the surface shown to the
right (Fig.~1(b)).

The closed string is represented by the puncture
at $C$, where we have a conical singularity of finite area. The surface
is flat everywhere except at this point, where the curvature is positive
and there is a defect angle of $\pi$. It corresponds to a first order
pole of a quadratic differential. Let us explain this. The metric $\rho$
on the string diagram (recall $dl = \rho |dz|$) arises from a quadratic
differential $\varphi = \phi(z) dz^2$ via the relation
$\rho = |\phi |^{1/2}$. For our case, in the natural coordinate $w$,
where the strip appears as shown in Fig.~2, $\rho (w) =1$
and $\varphi (w) = dw^2$. The $w$ coordinate, centered on $C$, is
mapped to a full neighborhood of the complex plane via $z=w^2$ (Fig.~2(b)).
Using $\phi (z) dz^2 = \phi (w) dw^2$, one readily finds that
$\varphi(z) = dz^2 /z$, hence the name first order pole of a
quadratic differential. The horizontal trajectories of the quadratic
differential are shown both in the $w$ and in the $z$ plane (for more
details see [\strebel ]).

This vertex can define an off-shell amplitude. The strip defines
canonical coordinates as far as the open string puncture is concerned.
For the closed string puncture the $z$ coordinates can be defined
to be the local coordinates. Namely, for any point $P$ sufficiently
close to $C$ we define $z(P) = (w(P))^2$. This allows us to insert
off-shell closed string states at $C$. The $z$ coordinates do not
extend beyond the images of the vertical boundaries of the strip.
The coordinates cannot be extended either beyond the image of a horizontal
$w$ line where a curvature singularity appears.
This difficulty becomes relevant for more complicated string diagrams.
\medskip
\noindent
$\underline{\hbox{Open-Open-Closed Amplitude}}.\,$ Analysis of this
amplitude will now show that, in contrast with the case of light-cone
field theory, or the open-closed covariant theory, there will
be no need for an elementary interaction in order to cover fully
the moduli space of the relevant Riemann surfaces. These surfaces are
disks with one closed string puncture and two open string punctures
in the boundary of the disk, and have one modular parameter, which
can be chosen to be the angle $\theta$ shown in Fig.~3(a). Using our
Feynman rules we obtain the string diagram shown to the right. The closed
string, represented by the puncture at $P$, via the open-closed vertex
turns instantaneously into an open string, which propagates for time
$T$, and splits, via the symmetric open string vertex into open strings
$1$ and $2$. Note that large $T$ corresponds to small $\theta$,
since effectively the open strings are becoming close to each other.
When $T\to 0$ the
string diagram turns into a completely flat semiinfinite strip
with the closed string inserted on the middle of the strip (with
no curvature singularity). This happens because at $T\to 0$, segments
$OA$ and $OB$ get identified. This configuration corresponds to
$\theta = \pi /2$. Since we are dealing with orientable open strings,
configurations with $\theta$ and $\pi -\theta$ are not equivalent.
The diagram in Fig.~3(a) gave us the region $\theta \in [0,\pi/2]$,
the diagram in Fig.~3(b) gives us the region $\theta \in [\pi /2 , \pi]$.
In the second diagram, when $T\to 0$, we find exactly the same
configuration as we did in the previous case. It is clear we have
covered completely moduli space, and we did not need an extra
elementary interaction. As punctured surfaces with metrics the
two string diagrams obtained in the limit as $T \to 0$ are {\it identical.}

Covering moduli space, however, is not sufficient in order to have
a well defined off-shell amplitude. We do not seem to be able to get
a consistent BRST invariant amplitude if we
insist on having off-shell closed strings. The problem arises because
we are unable to define the local coordinates around the closed
string puncture smoothly over moduli space. For any specific surface,
it is sufficient to specify coordinates in any nonvanishing neighborhood
of the puncture. Note that, given the way we are defining coordinates
around the closed string puncture, they cannot be extended
beyond $|z| = T^2$, since the flat strip where the closed string
puncture lies has a metric singularity at $w = iT$, the
interaction point of the three open strings.
As we let $T \to 0$, we eventually cannot define local coordinates
around the closed string punctures. For any $T \not= 0$ we can
define coordinates, but the limit as $T$ goes to zero does not
exist. Indeed the closed string puncture, which sits on top of
a first order pole of the quadratic differential collides with
the first order zero at the interaction point. The result is a
regular point, and we lose the ability to define coordinates
smoothly. This discontinuity of local coordinates on the boundary
of the two string diagrams implies that we cannot match the off-shell
amplitudes, and prevents us from getting a BRST invariant amplitude.

Let us consider now the amplitude for scattering of three open
strings and one closed string via a disk. We just check that
the Feynman rules we have derived so far suggest that we cover
fully moduli space. We do so by verifying that the boundaries
of the Feynman graphs match.  The Feynman rules are so simple,
that there is just one type of
graph structure involving two three-open-string vertices
and one open-closed vertex. A particular assignment of the
open strings is shown in Fig.~4 (a).  The limits when either
$T$ or $T'$ go to infinity correspond to open string poles,
and are therefore not considered to be boundaries of moduli
space. We only need to consider the cases when $T$ or $T'$ go
to zero. When $T$ goes to zero the diagram matches
with the diagram of Fig.~4(b). When $T'$ goes to zero, it
matches with the diagram of Fig.~4(c). Writing out all the
relevant assignments of open strings, similar arguments show
that all boundaries match. Instead of checking further examples,
we will now use the geometrical recursion relations of the theory
to argue systematically that further elementary interactions
are not needed in order to cover moduli space.
\medskip
\noindent
$\underline{\hbox{Satisfying the Recursion Relations}}.\,$ We have
now evidence that the Feynman rules of the following Lagrangian
$$\L = \int ( AQA + {2\over 3}g A^3) + \hbar^{1\over 2} g \int
A \Psi ,$$
generate string diagrams that provide a single cover of moduli
space. The first term in this lagrangian is the Witten open string
field theory, with $A$ the open string field, and the second term
is the open-closed interaction that we have discussed above.
As explained in [\zwiebach ], for consistency of the loop expansion,
this interaction appears at order $\hbar^{1/2}$. Note that there is no
kinetic term for the closed string field $\Psi$, but the Feynman
graphs will give rise to all required closed string poles.
This action does not enable us to compute amplitudes involving
surfaces without boundaries, namely, pure closed string processes.

The above lagrangian may not to be truly fundamental, since
it fails to give BRST invariant off-shell amplitudes for procesess
involving off-shell closed strings. It generates, however, all the
relevant moduli spaces in an
extremely simple way.

Since we expect covering of moduli space by the string diagrams
generated by the Feynman rules, a geometrical consistency condition
relating vertices must hold. This condition, first derived for
closed strings in [\sonodazwiebach ], was generalized to a theory including
open strings in [\zwiebach ]. For each scattering amplitude, which involves
a specific moduli space, there is a potential elementary vertex filling some
region of that moduli space. The consistency condition demands
that the surfaces in the boundary of the region of moduli space
filled by this vertex should coincide with the appropriate
string diagrams involving one propagator only, in the limit when
the propagator collapses.
The moduli space, or type of amplitude is specified by the genus
$G$ of the surface, the number $N$ of closed string punctures, the
number $B$ of boundary components, and the numbers $m_i, i=1\cdots B$,
which give the number of open string punctures in each boundary
components. A particular vertex will be denoted by
$\V^{G,N}_{B,M}$ (with $M=\sum m_i$). The dimension of a vertex
will denote the dimension
of the corresponding moduli space. Given that we only have
open string propagators, the
consistency condition is that given in Fig.~5, a subcase
of the consistency condition given in [\zwiebach ], where we also had
closed string propagators. (A discussion of geometrical consistency
conditions can be found in [\zwiebachtalk ].)

We want to see if it is consistent to have just a three open string
vertex $\V^{0,0}_{1,3}$ and an open-closed vertex $\V^{0,1}_{1,1}$,
and to set all others $\V^{G,N}_{B,M}$ to zero.
There is no need to check
an infinite number of vertices. The consistency conditions relate
a vertex of dimension $d$, in the left hand side of the equation
to either two vertices whose dimensions add up to $(d-1)$, or to
a single vertex of dimension $(d-1)$. The two vertices we have
correspond to $d=0$ and we claim that all other $d=0$ vertices
vanish. If we can show that under this assumption no $d=1$
vertex is necessary, then it is consistent to set all higher
dimensional vertices ($d \geq 2$) equal to zero.

The dimension zero vertices we include correspond to
the disk with three open string punctures and the disk with
one open and one closed puncture (we do not include the
disk with one or two open string punctures, nor the
disk with one closed string puncture).
Since the consistency
condition is a consequence of correct covering of moduli space,
in order to verify that the dimension one vertices can be set
to zero we check that these moduli spaces are completely
covered once we use the Feynman rules based on our dimension zero
vertices.

The dimension one moduli spaces are: (a) the disk with four open
string punctures, which is well covered in the usual open string
theory [\witten ], and therefore does not require the introduction of an
elementary vertex; (b) the disk with one closed string
and two open strings, which is also well covered, as we found
above that no open-open-closed vertex was required;
(c) the annulus with one open string puncture, which is the open
string tadpole calculated with Witten's vertex, and which covers
moduli space (but gives a singular Riemann surface in the right
hand side of the geometrical equation, leading to regulatization
problems), and finally, (d) the disk with two closed string
punctures; this moduli space is well
covered by the relevant string diagram, shown in Fig.~6 (this
configuration also gives a singular term in the geometrical
equation, corresponding to $T=0$). Since we have found complete
covering, this shows that all dimension one vertices can be set to zero,
and as a consequence all higher dimension vertices can be set
to zero as well.

The consistency condition that we have verified is a necessary
condition for getting a single cover of moduli space. Since it is
not a sufficient condition, rigorously speaking, we have only
obtained evidence that we have a single cover. The evidence will
be furthered now by
showing that the Feynman rules build minimal area string diagrams.
A complete proof would require an analysis
similar to that of [\zwiebachcmp ], namely establishing an existence and
uniqueness theorem for quadratic differentials having first and
second order poles. The methods of [\zwiebachcmp] are likely to be
applicable, but we will not try to construct such a proof here.
\medskip
\noindent
$\underline{\hbox{Minimal Area Property of the String Diagrams}}.\,$
It was explained precisely in [\zwiebachcmpnext] why the minimal
area property guarantees that
no surface is produced more than once by the Feynman rules. The basic idea
is that two different Feynman diagrams must produce different
metrics, both of minimal area. But for a given surface there is just
one metric of minimal area, thus the two Feynman diagrams must
be building different Riemann surfaces.

The minimal area property for the present case follows from the
following two
facts: (a) on every strip the open curves extending from one boundary
to the other are nontrivial, and (b) any nontrivial open
curve is longer or equal to $\pi$. Property (a) holds because
if we double the string diagram we get a meromorphic Jenkins-Strebel
quadratic differential (with both first order and second order poles).
In this surface the double of the open curves lying on a strip are
curves homotopic to core curves of the ring domains. Since
core curves are nontrivial, the closed curves in question must be
nontrivial. But a nontrivial closed curve built by
doubling must arise from a nontrivial open curve [\zwiebachcmpnext ].
Property (b) holds because any
nontrivial open curve must cross the line that goes along the middle of all
the strips in the string diagram. Therefore,
the metric satisfies the necessary length conditions.
The area is minimal because it is minimal under the weaker condition that
the open curves in each of the strips be greater or equal to
$\pi$. This is so because it is made of flat strips.
\medskip
\noindent
$\underline{\hbox{Interpolating String Field Theories}}.\,$
We have seen that it is possible to have closed string punctures
and to have length conditions on open curves only. This is the
limit case of a more general problem that imposes length
conditions on both open and closed curves.
We can pose the following problem for open-closed string theory:
{\it the string diagrams are defined by the metric of
minimal area under the condition that all nontrivial Jordan open curves be
longer or equal to $l_o$ and all nontrivial Jordan closed curves be longer
or equal to $l_c$}. As long as $l_o \not= 0$ we can scale the string
diagrams to $l_o = \pi$, and we get a one parameter family of
string diagrams. They interpolate from the open-closed field theory
[\zwiebach ],
which has $l_c = 2\pi$, to the theory studied here, which corresponds
to $l_c = 0$. A sequence of open closed string diagrams is shown in
Fig.~7. As long as we keep $l_c \not= 0$
the string diagrams may be reconstructed using vertices and
propagators, in a way similar to that of [\zwiebach ].
Thus we get a a one-parameter family of consistent off-shell string field
theories. Some of the minimal area metrics will not arise from quadratic
differentials. As $l_c \rightarrow 0$ one can verify that the higher
open-closed vertices fill progressively smaller regions of moduli
space.
For $l_c =0$, only the open-closed vertex is relevant, and
the off-shell property is lost. We then get simple string
diagrams arising from quadratic
differentials, as we showed earlier.

For closed strings one can also generalize the string diagrams.
Since modular transformations do not mix Jordan closed curves
homotopic to punctures with Jordan closed curves that are not
homotopic to a puncture we can put different length conditions
for the two types of curves. We obtain a one parameter family
of generalized closed string
diagrams by requiring that {\it the area be minimal under the
conditions that nontrivial Jordan closed curves not homotopic to punctures
be longer or equal to $l_c$ and those homotopic to punctures
longer or equal to $l_p$}. The standard closed string field theory
corresponds to $l_c = l_p = 2\pi$. Another familiar case is
$l_c = 0$ and $l_p=2\pi$; it corresponds to the Strebel
construction of every surface as a contact interaction [\saadizwiebach ].
Thus the one parameter family of string diagrams with $l_p=2\pi$
and $l_c$ variable interpolate from the standard closed string diagrams
into the Strebel construction of every surface as a polyhedron.

The string diagrams where both $l_c$ and $l_p$ are different from
each other, and different from zero do not seem to arise from
naive Feynman rules of string field theory.
This is the case because putting different length
conditions for curves surrounding punctures and for other curves
is incompatible with naive sewing. If we sew together two elementary
vertices we get an internal cylinder with circumference $l_p$.
If $l_p < l_c$, this violates the length conditions, and if
$l_p > l_c$, it is clear that for a sufficiently long cylinder,
the saturation condition [\zwiebachqcs ] is violated, and the metric
could not be of minimal area (in the middle of the tube we could not
find a nontrivial curve of length $l_c$). In these string diagrams all
degenerations happen nicely via long tubes, but we have no
off-shell factorization.

The case $l_c = 2\pi$ and $l_p=0$, corresponds to no
conditions on curves homotopic to the punctures, thus the punctures
appear typically as conical singularities.
Note, however, that in contrast with the case of open strings,
we do not expect off-shell problems here,
since the closed string punctures cannot collide (any nontrivial
curve surrounding two punctures must be longer or equal to $2\pi$).
It may be interesting to elucidate the structure of these string
diagrams.

We hope to have illustrated here the power of minimal area methods.
All decompositions of moduli space used in covariant string
field theory arise from minimal area metrics. We have found here
that we can easily generate families of decompositions that
interpolate between various known and new decompositions.
In one case we get a family of string field theories. A possible
utility of our results would be an enlarged framework for string
field theory where background independence would be easier to
understand.
\ack
I am grateful to R. Dijkgraaf, T. Kugo and E. Witten for discussions.
\medskip
\refout
\endpage
\noindent
{\bf Figure Captions}
\medskip
\noindent
Figure (1). (a) Finding out where to place
the closed string puncture on the strip.
Positions $A$ and $B$ are not possible. The closed string must be
placed at $C$ and
the segments $CD$ and $CD'$ have to be identified. (b) The resulting
vertex geometry.
\medskip
\noindent
Figure (2). (a) The open-closed vertex shown in the $w$-plane, where
it appears as a semiinfinite strip with the closed string puncture
at $C$, and with $CD$ and $CD'$ identified. (b) The mapping $z=w^2$
maps the strip into the $z$ plane. Here the quadratic differential
takes the form $dz^2/z$ corresponding to a first order pole.
The horizontal trajectories are shown both in the $w$ and in the
$z$ plane.
\medskip
\noindent
Figure (3). In open-open-closed scattering the
relevant moduli space can be parametrized by the angle $2\theta$
separating the open string punctures in a unit disc where the
closed string puncture is at the center. (a) $\theta \in [0, \pi /2 ]$;
the string diagram having an open string propagator, and
the limit as this propagator collapses.
(b) $\theta \in [\pi /2 , \pi ]$.
It is clear that in the limit when the two propagators collapse
the resulting surfaces (with metrics) are identical.
\medskip
\noindent
Figure (4). String diagrams for scattering of
three open strings and one closed string via a disk.
Configurations match as the open string propagators collapse.
\medskip
\noindent
Figure (5). The geometrical consistency conditions for elementary
vertices in the extended covariant open string theory. The boundary
of an elementary interaction must coincide with a configuration
with a single open string propagator, in the limit when this
propagator collapses. The wiggly lines correspond to external
closed strings, and the other lines represent open strings.
The heavy dots represent boundary components.
\medskip
\noindent
Figure (6). The string diagram corresponding to the scattering
of two closed strings via a disk. The corresponding moduli space
is one dimensional and is generated completely by the string
diagram shown above. When the open string propagator collapses
the two closed string punctures collide giving a singular limit.
\medskip
\noindent
Figure (7). We show how the open-closed vertex varies as we
change the length condition on the closed curves. We show
four cases, corresponding to $l_c > l_o$, $\, l_c = l_o$,
$\, l_c < l_o$, and $l_c = 0$.
\end